\documentclass{PoS}

\usepackage{pjournal}

\title{Dust in High-Velocity Clouds : relevance for Planck}

\ShortTitle{Dust in High-Velocity Clouds : relevance for Planck}

\author{\speaker{Marc-Antoine Miville-Desch\^enes}%
         \thanks{A footnote may follow.}\\
        Institut d'Astrophysique Spatiale, Orsay, France\\
        E-mail: \email{mamd@ias.u-psud.fr}}

\author{Fran\c{c}ois Boulanger\\
        Institut d'Astrophysique Spatiale, Orsay, France\\
        E-mail: boulanger@ias.u-psud.fr}

\author{Peter G. Martin\\
        Canadian Institute for Theoretical Astrophysics, Toronto, Canada\\
        E-mail: pgmartin@cita.utoronto.ca}

\author{Felix Jay Lockman\\
  NRAO, Green Bank, WV, USA\\
  E-mail: jlockman@nrao.edu}

\author{William T. Reach\\
  Spitzer Science Center, Pasadena, CA, USA\\
  E-mail: reach@ipac.caltech.edu}

\author{Alberto Noriega-Crespo\\
  Spitzer Science Center, Pasadena, CA, USA\\
  E-mail: alberto@ipac.caltech.edu}

\abstract{The recent detection of dust emission in Complex C, the largest 
High-Velocity Cloud (HVC) on the sky,
opens a very interesting perspective for Planck. The HVC dust temperature 
determined using IRAS and Spitzer
observations is low ($T\sim10.7$~K) in accordance with its great distance
from the Galaxy ($> 5$~kpc). 
Peak column density in 30 arcmin beam is $N_H \sim 10^{20}$~cm$^{-2}$ which is
typical of HVCs and similar to cirrus column density in diffuse regions.
On the other hand HVCs appear to be very clumpy at smaller angular scales; several
observations at the arcminute scale resolution show significant structure
and higher brightness contrasts than in typical cirrus emission.
In this contribution we show that, even with their moderate column density,
the cold temperature, high emissivity and high column density contrast of 
HVCs should lead to significant and detectable emission in the Planck-HFI frequency range.
In order to separate the HVC emission from the Galactic cirrus emission,
the use of 21 cm observations will be mandatory.

}

\FullConference{CMB and Physics of the Early Universe\\
                 20-22 April 2006\\
                 Ischia, Italy}

\begin{document}

\section{Introduction}

\begin{figure}
\includegraphics[width=\linewidth, draft=false]{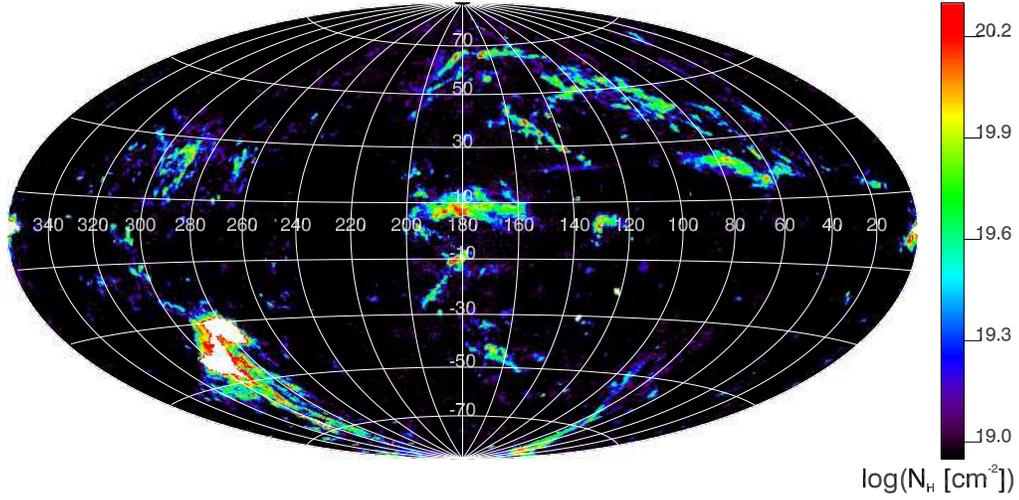}
\caption{All-sky map of the High-Velocity Clouds (HVC) at 21 cm as given
by the Leiden-Argentine-Bonn (LAB) survey \cite{kalberla2005}. 
The bright region at (l$\sim$290, b$\sim$-40) is the Magellanic Clouds.}
\label{fig_hvc_allsky}
\end{figure}

High-Velocity clouds (HVC) are diffuse emission structures first detected
at 21 cm  which have average velocity forbidden by the Galactic rotation. 
HVCs are observed on 30\% of the sky (see Fig.~\ref{fig_hvc_allsky}) 
and have angular sizes ranging from less 
than one square degree (Compact HVC) to 2000 square degrees for Complex C.
The origin of HVCs is still a matter of debate; they could be
1) gas expelled from the Galactic disk
due to star formation activity (i.e. Galactic fountain), 2) interaction with satellite
galaxies - including the Magellanic clouds, 3) condensation of the coronal gas
and 4) accretion of primordial gas - gas still in the Local group cluster that
has never formed star. To conclude on the origin of HVCs, measurements of 
their distance (and therefore mass) as well as their metallicities are essential. 

Several arguments are invoked 
to exclude the Galactic fountain scenario, including the fact that HVCs 
do not follow Galactic rotation and that
several HVCs are too massive to have been expelled even by several supernovae;
when distance estimates are available the mass of HVCs are in the $10^5-10^6$~M$_\odot$ range. 
HVCs have also been observed around other galaxies like M31 
\cite{thilker2004}
where they are at distances of $<60$~kpc from the disk and 
have masses ranging from $10^5$ to $10^7$~M$_\odot$.

It is now widely considered that many of them 
might be infalling clouds fueling the Galaxy with low metallicity gas. 
This hypothesis received strong support from ultraviolet observations showing that some 
HVCs have a subsolar metallicity \cite{wakker99} and a D/H ratio compatible with primordial 
abundances of deuterium \cite{sembach2004}. 
There is also the suggestion \cite{blitz99} that Compact-HVCs could
be the residual of the formation of the Local group and the equivalent of the 
mini-dark halos which are observed in simulations of structure formation. 
In this scenario the largest HVCs on the sky would be some of those clouds that are 
falling on the Galaxy, trapped by its gravitational potential well.

The infall of fresh and metal-poor gas on the Galaxy is required to explain several
important properties of Galactic ecology : the star formation
rate, the constant metallicity of G stars and the presence of a bar. 
The exact nature of this infalling gas is still unknown. 
Some of it could be the HI seen in HVC but looking
at external galaxies like M31, 10 to 100 more mass than what is seen in HI is needed.
Part of this mass is most probably in the form ionized gas that continuously condense
into HI but there is most likely also dark matter. 
Anyhow HVCs could provide essential clues in our understanding of
galaxy formation and evolution.

\section{First detection of dust emission in a HVC}

Recently we \cite{miville-deschenes2005} presented the first detection of dust emission
from a HVC made by comparing sensitive Spitzer Space Telescope infrared and 
Green Bank Telescope 21 cm observations.
This study focused on the Spitzer Extra-Galactic First Look Survey (XFLS) field,
a diffuse HI area at high Galactic latitude, located on 
the edge of Complex C (subsolar metallicity 
($\sim 0.1-0.3$) and distance greater than 5 kpc from the sun \cite{wakker2001}). 
Previously dust emission has been unsuccesfully looked for in HVCs using IRAS data 
\cite{wakker86}.
Upper limits were found to be compatible with their low HI column densities, 
low metallicity (and therefore low dust/gas ratio) and their large distance 
(i.e. fainter radiation field). The detection of dust emission \cite{miville-deschenes2005}
has only been made possible by the use of more sensitive infrared data (Spitzer and IRIS - 
a reprocessing of the IRAS data \cite{miville-deschenes2005a})
and higher resolution and more sensitive 21 cm observations (GBT).

Dust in this HVC is found to be significantly colder than in the local interstellar medium, 
even at the 3$\sigma$ limit ($T_{BG}^{HVC} = 10.7_{-3.0}^{+2.9}$),  
which is consistant with a lower radiation field than in the Solar neighborhood 
due to its distance to the Galaxy \cite{wakker86}.
In addition, from the correlation between infrared and 21 cm data, we report 
a far-infrared emissivity $\epsilon_{HVC}$ (i.e. emission from big grains per H atom)
higher in the HVC than in the Solar neighborhood ($\epsilon_{HVC}/\epsilon_{soln} > 1.6$ (3$\sigma$)). 
For dust properties typical of the local interstellar medium, 
$\epsilon_{HVC}$ should be proportional to the dust-to-gas mass ratio which, for standard metal 
depletion on grains, would scale with the metallicity. 
The fact that $\epsilon_{HVC}/\epsilon_{soln}$ is greater than the metallicity 
in Complex C ($ Z_{HVC}/Z_\odot = 0.2\pm0.1$) opens interesting questions about
the nature of the baryons in HVCs (see \cite{miville-deschenes2005} for details).

\section{Can HVCs be detected with Planck ?}

\begin{figure}
\begin{center}
\includegraphics[width=10cm, draft=false]{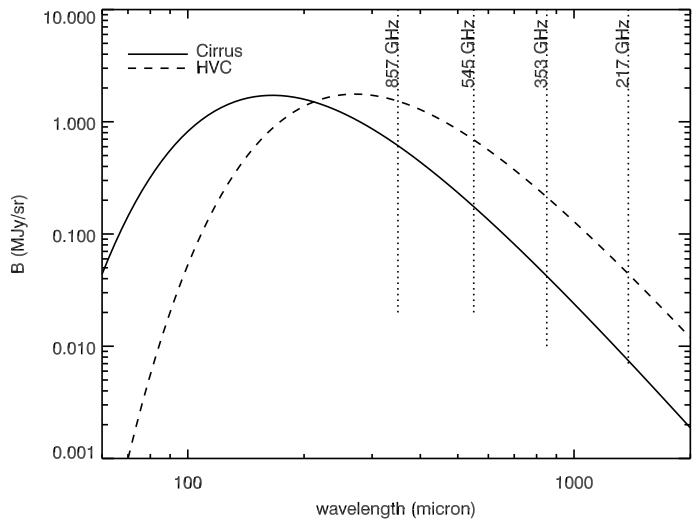}
\caption{Comparison of thermal dust emission for a typical cirrus ($T_{dust}=17.5$~K)
and a HVC ($T_{dust}=10.7$~K) assuming a dust spectral index $\beta=2$. 
The vertical dotted lines show the expected brightness sensitivity for
the four highest frequencies of Planck-HFI ($1\sigma$ for 14 months in 5 arcmin pixel
is 0.02, 0.02, 0.01 and 0.007 MJy/sr at 857, 545, 353 and 217 GHz).}
\label{fig_Dust_HVC_spectrum}
\end{center}
\end{figure}

The detection of dust emission in Complex C
shows that dust in HVCs is cold ($T_{dust}\sim10$~K) and 
has a higher emissivity per $N_{HI}$ than
in Galactic cirrus. Using the Complex C numbers we estimated the emission of HVCs
in the four highest frequency bands of HFI. In Fig.~\ref{fig_Dust_HVC_spectrum} we show the
big grain thermal emission for a HI column density of $N_{HI}=10^{20}$~cm$^{-2}$
for a cirrus and a HVC. 
To make a first order comparison in the Planck-HFI wavelength range 
between cirrus and HVC emission (for which nothing
is known about the dust spectrum) we used a very simple dust model with a 
dust spectral index $\beta=2$.
According to this simple model, dust emission from HVCs could dominate
the thermal dust emission in low cirrus column density regions. This leaves
no doubt about the capacity of Planck-HFI to detect the HVC thermal dust emission.

At scales larger than 30 arcmin the ratio of HVC to cirrus HI column density is $<$0.5. 
On the other hand observations at higher angular resolution of
specific HVCs have shown that this ratio increases significantly
at smaller scales. The structure of HVCs is generally more clumpy and has stronger
contrast than the cirrus emission. This difference in morphology 
should help identify and separate the HVC from the local cirrus emission in the Planck-HFI data.


\begin{thebibliography}{99}

\bibitem{blitz99}
Blitz, L., Spergel, D.~N., Teuben, P.~J., Hartmann, D. \& Burton, W.~B.
\emph{High-Velocity Clouds: Building Blocks of the Local Group,}
\newblock 1999, \ApJ, 514, 818.

\bibitem{kalberla2005}
{Kalberla}, P. M.~W., et al. 
\emph{The Leiden/Argentine/Bonn (LAB) Survey of Galactic HI. Final data release of the combined 
  LDS and IAR surveys with improved stray-radiation corrections,}
\newblock 2005, \AaA, 440, 775.

\bibitem{miville-deschenes2005}
Miville-Desch\^enes, M.~A., Boulanger, F., Reach, W.~T. \& Noriega-Crespo, A.
\emph{The First Detection of Dust Emission in a High-Velocity Cloud,}
\newblock 2005, \ApJ, 631, L57--L60.

\bibitem{miville-deschenes2005a}
Miville-Desch\^enes, M.~A. \& Lagache, G.
\emph{IRIS: A New Generation of IRAS Maps}
2005, ApJS, 157, 302

\bibitem{sembach2004}
Sembach, K.~R., et al.
\emph{The Deuterium-to-Hydrogen Ratio in a Low-Metallicity Cloud Falling onto the Milky Way,}
\newblock 2004, ApJS, 150, 387.

\bibitem{thilker2004}
Thilker, D.~A., et al.
\emph{On the Continuing Formation of the Andromeda Galaxy: Detection of H I Clouds in the M31 
  Halo,}
\newblock 2004, \ApJ, 601, L39.

\bibitem{wakker86}
Wakker, B.~P. \& Boulanger, F.
\emph{A search for dust in high-velocity clouds,}
\newblock 1986, \AaA, 170, 84.

\bibitem{wakker99}
Wakker, B.~P., et al.
\emph{Accretion of low-metallicity gas by the Milky Way,}
\newblock 1999, \Natur, 402, 388.

\bibitem{wakker2001}
Wakker, B.~P.
\emph{Distances and Metallicities of High- and Intermediate-Velocity Clouds,}
\newblock 2001, ApJS, 136, 463.

\end{thebibliography}
\end{document}